\documentstyle[preprint,aps,epsf]{revtex}
\begin{document}
\draft

\title{Inertial Effects on Fluid Flow through Disordered Porous Media}

\author{J. S. Andrade Jr.$^{1,3}$, U. M. S. Costa$^{1}$, M. P. Almeida$^{1}$,
H. A. Makse$^{2}$, and H. E. Stanley$^{3}$}

\address{$^1$Departamento de F\'{\i}sica, Universidade Federal do Cear\'a,
60451-970 Fortaleza, Cear\'a, Brazil \\
$^2$ Schlumberger-Doll Research, Old Quarry Road, Ridgefield, CT 06877 \\
$^3$ Center for Polymer Studies and Physics Dept., Boston University,
Boston, MA 02215}

\maketitle
\begin{abstract}
We investigate the origin of the deviations from the classical Darcy
law by numerical simulation of the Navier-Stokes equations in
two-dimensional disordered porous media. We apply the Forchheimer
equation as a phenomenological model to correlate the variations of
the friction factor for different porosities and flow conditions.  At
sufficiently high Reynolds numbers, when inertia becomes relevant, we
observe a transition from linear to non-linear behavior which is
typical of experiments. We find that such a transition can be
understood and statistically characterized in terms of the spatial
distribution of kinetic energy in the system.
\end{abstract}

\pacs{Phys. Rev. Lett. {\bf  82}, 5249 (1999)}


\narrowtext

\newpage

A standard approach in the investigation of single-phase fluid flow in
microscopically disordered and macroscopically homogeneous porous
media is to characterize the system in terms of Darcy's law
\cite{Dul79,Sah94,Adl92}, which assumes that a {\it global\/} index,
the permeability $k$, relates the average fluid velocity $V$ through
the pores with the pressure drop $\Delta P$ measured across the
system,
\begin{equation}
V = -{k \over \mu}{\Delta P \over L}.
\label{eq1}
\end{equation}
Here $L$ is the length of the sample in the flow direction and $\mu$
is the viscosity of the fluid. However, in order to understand the
interplay between porous structure and fluid flow, it is necessary to
examine {\it local} aspects of the pore space morphology and relate
them to the relevant mechanisms of momentum transfer (viscous and
inertial forces). This has been accomplished in previous studies
\cite{Can90,Kos92,Mar92,Sch93,Mar94,Kop97,And95} where computational
simulations based on a detailed description of the pore space have
been quite successful in predicting permeability coefficients and
validating well-known relations on real porous materials.

In spite of its great applicability, the concept of permeability as a
global index for flow, should be restricted to viscous flow conditions
or, more precisely, to small values of the Reynolds number.  Unlike
the sudden transition from laminar to turbulent flow in pipes and
channels where there is a critical Reynolds number value separating
the two regimes, experimental studies on flow through porous media
have shown that the passage from linear (Darcy's law) to nonlinear
behavior is more likely to be gradual (see Dullien \cite{Dul79} and
references therein).  It has then been argued \cite{Dul79} and
confirmed by numerical simulations \cite{Edw90,Koc97} that the
contribution of inertia to the flow in the pore space should also be
examined in the framework of the laminar flow regime before assuming
that fully developed turbulence effects are present and relevant to
momentum transport. Here we show by direct simulation of the
Navier-Stokes equations that the departure from Darcy's law in flow
through high porosity percolation structures ($\epsilon>{\epsilon}_c$,
when ${\epsilon}_c$ is the critical percolation porosity) and at
sufficiently high Reynolds numbers can also be explained in terms of
the inertial contribution to the {\it laminar} fluid flow through the
void space. The calculations we perform do not apply for unstable or
turbulent Reynolds conditions. We then demonstrate that it is possible
to {\it statistically} characterize the transition from linear to
nonlinear behavior in terms of the distribution of kinetic
energy. This allows us to elucidate certain features of the fluid flow
phenomenon in irregular geometries that have not been studied before.

Our model for the pore connectivity is based on the general picture of
site percolation disorder. Square obstacles are randomly removed from
a 64x64 square lattice until a porous space with a prescribed void
fraction $\epsilon$ is generated. The mathematical description for the
detailed fluid mechanics in the interstitial pore space is based on
the assumptions that we have steady state flow in isothermal
conditions and the fluid is continuum, Newtonian and
incompressible. Thus, the continuity and Navier-Stokes equations
reduce to
\begin{equation}
{\bf{\nabla\cdot u}}=0~~,
\label{eq2}
\end{equation}
\begin{equation}
\rho~{\bf{u\cdot\nabla u}} = -{\nabla p} +
\mu~{\bf{\nabla}}^{2}{\bf{u}}~~,
\label{eq3}
\end{equation}
where $\rho$ is the density of the fluid and $\bf u$ and $p$ are the
local velocity and pressure fields, respectively. We use the nonslip
boundary condition at the whole of the solid-fluid interface. End
effects of the flow field established inside the pore structure
(particularly significant at high Reynolds conditions) are minimized
by attaching an inlet and an outlet to two opposite faces. At the
inlet a constant inflow velocity in the normal direction to the
boundary is specified, whereas at the outlet the rate of velocity
change is assumed to be zero (gradientless boundary
condition). Instead of periodic boundary conditions, we close the
remaining two faces of the system with two additional columns of
obstacles. This insulating condition reproduces more closely the
experimental setup usually adopted with real rocks and permeameters.
The Reynolds number is defined here as $Re \equiv \rho V d_p /\mu$
where $d_p$ is the grain diameter \cite{grain}. For a given
realization of the pore geometry and a fixed $Re$, the local velocity
and pressure fields in the fluid phase are numerically obtained
through discretization (see \cite{And95} for numerical details) by
means of the control volume finite-difference technique
\cite{mesh,Pat80}. Finally, from the area-averaged pressures at the
inlet and outlet positions, the overall pressure drop can be readily
calculated.

The classical approach to macroscopically characterize the effect of
inertia on flow through real porous media is to use the Forchheimer
equation \cite{Dul79,Sah94},
\begin{equation}
-{\Delta P \over L} = {\alpha \mu V} + {\beta \rho {V^2}}.
\label{eq4}
\end{equation}
The coefficient $\alpha$ corresponds to the reciprocal permeability of
the porous material and $\beta$ is usually called the ``inertial
parameter''.  Both $\alpha$ and $\beta$ should depend on the porosity
$\epsilon$ of the porous material. At sufficiently low velocities,
Eq.~(4) reduces to Darcy's law, Eq.~(1). The term $\beta \rho {V^2}$
can be interpreted as a second order correction to account for the
contribution of inertial forces in fluid flow. Equation~(4) is not a
purely empirical expression, since it can be derived by an appropriate
average of the Navier-Stokes equation for one-dimensional, steady
incompressible laminar flow of a Newtonian fluid in a rigid porous
medium \cite{Dul79}. Rearranging (\ref{eq4}) in the form,
\begin{equation}
f = {1 \over Re'} + 1~~,
\label{eq5}
\end{equation}
where $f\equiv-\Delta P / {L \beta \rho {V^2}}$ and $Re'\equiv\beta
\rho V / {\alpha \mu}$, we obtain a {\it friction factor-Reynolds
number} type of correlation which is presumably ``universal.''
Equation~(5) has been extensively and successfully used to correlate
experimental data from a large variety of porous materials and a broad
range of flow conditions \cite{Dul79}. Certainly, a better
representation for experimental data in the non-Darcy flow regime can
be obtained with the addition to the Forchheimer equation of third
order corrections in the velocity \cite{Sah94,Edw90,Koc97}. The
theoretical basis for this type of correction term, however, is still
controversial.

Figure~1 shows the results of our flow simulations in terms of the
Forchheimer variables $f$ and $Re'$ for three different values of
lattice porosity ($\epsilon=0.7$, $0.8$ and $0.9$). After computing
and averaging the overall pressure drops for all realizations at
different values of $\epsilon$ and $Re$, we fit the results with
Eq.~(4) to estimate the coefficients $\alpha$ and $\beta$ and
calculate $f$ and $Re'$. In agreement with real flow experiments, we
observe a transition from linear (Darcy's law) to nonlinear flow.
Moreover, the point of departure from linear to nonlinear behavior in
the range ${10^{-2}}<Re'<{10^{-1}}$ is consistent with previous
experimental observations. However, in the region $6.2\times
10^{-2}<Re'<1.8$, the Forchheimer equation generally overestimates the
computed values of the friction fraction. In addition, for a fixed
value of $Re'$, the variability in this transition region is
sufficient to suggest a dependence of the type $f=f(Re',\epsilon)$
\cite{Lee97}.

The flow distribution in two-dimensional incompressible systems can be
conveniently described in terms of the stream function $\psi$
\cite{stream}.  Figure~2a shows the contour plot of $\psi$ for a
typical realization of a highly porous void space ($\epsilon=0.9$)
subjected to low Reynolds conditions, $Re=0.0156$. In spite of the
well-connected pathways available for flow at this large porosity
value, the predominant viscous forces in the momentum transport
through the complex void geometry generates well defined
``preferential channels'' of fluid flow. As shown in Fig.~2b, the
situation is quite different at high $Re$, where the degree of
channeling is clearly less intense than in Fig.~2a. In the case of
Fig.~2b, due to the relevant contribution of inertial forces to the
flow, the distribution of streamlines along the direction orthogonal
to the main flux $y$ becomes more homogeneous.

The channeling effect can be statistically quantified in terms of the
spatial distribution of kinetic energy in the flowing system. In
analogy with previous work on localization of vibrational modes in
harmonic chains \cite{Rus94}, we define a ``participation'' number
$\pi$,
\begin{equation}
\pi\equiv\left(n\sum_{i=1}^nq_i^2\right)^{-1}~~~~~~\left({1\over
n}\leq \pi \leq 1\right)~~,
\label{eq6}
\end{equation}
where $n$ is the total number of fluid cells in the numerical grid
enclosing the physical pore space, ${q_i}\equiv
{e_i}/{\sum_{j=1}^{n}{e_j}}$, ${e_i}\propto ({u_i}^2+{v_i}^2)$ is the
kinetic energy associated with each individual fluid cell, and $u_i$
and $v_i$ are the components of the velocity vector at cell $i$ in the
$x$ and $y$ directions, respectively.~From the definition Eq.~(6),
$\pi=1$ indicates a limiting state of equal partition of kinetic
energy (${q_i}=1/n$, $\forall i$). On the other hand, a sufficiently
large system ($n\rightarrow\infty$) exhibiting strong channeling
effects should correspond to a ``localized'' flow field, $\pi\approx
0$ \cite{fsc}.  We calculate the function $\pi$ for 10 pore space
realizations generated with $\epsilon=0.9$ at different $Re$. Due to
convergence difficulties and computational limitations on the
resolution of the numerical grid, we restrict our calculations to
$Re\leq 15.6$. As shown in Fig.~3, the participation number remains
constant, $\pi\approx 0.37$, for low $Re$ up to a transition point at
about $Re\approx 0.3$. Above this point, the flow becomes gradually
less localized ($\pi$ increases) as $Re$ increases. This transition
reflects the onset of inertial effects in the flow, and the
significant changes in $\pi$ above the transition point indicate the
sensitivity of the system to these nonlinearities. The large error
bars at low $Re$ indicate that $\pi$ is sensitive to structural
disorder if the viscous forces are effectively generating preferential
channels in the flow.

The difference between our results at low and high $Re$ can be better
understood if we remember that viscous effects extend a long way at
low Reynolds conditions, so that distant boundaries may have a large
effect on the streamlines.  It is then interesting to visualize the
distortions in the local velocity field when inertial forces become
important compared to viscous forces. In Fig.~4 we show the profiles
of the velocity magnitude at different positions along the main flow
direction $x$ in a typical realization of the porous media.  At low
$Re$ (Fig.~4a), the fluctuations in the velocity field are essentially
smooth in shape, with peaks that closely correspond to the variations
in the local porosity. At high $Re$ (Fig.~4b), the situation becomes
quite different. Due to inertia, the effect on the flow field of the
disorder in the local pore geometry tends to propagate further the
fluctuations in the $x$ direction. We can follow in Fig.~4b the
changes in shape of the velocity magnitude at different $x$
positions. If there is an available {\it straight} void space pathway
for fluid flow, a peak generated at a smaller $x$ can {\it persist}
further up in the next profiles located at larger $x$ values.  As a
consequence, the velocity profiles at large $Re$ are more {\it rough}
than those at low $Re$.

In summary, to characterize the influence of inertial forces on the
flow of a single fluid in porous structures, we demonstrate that
incipient deviations from Darcy's law observed in several experiments
can be modeled in the laminar regime of fluid flow, without including
turbulence effects. The results of our simulations agree with numerous
experimental data which display a gradual transition at high $Re$ from
linear to nonlinear flow in the pore space.  Moreover, we show that
this flow transition can be characterized in terms of the partition of
kinetic energy in the fluid phase.  Namely, the flow at low $Re$ is
more ``localized'' due to channeling effects than the flow at high
$Re$ conditions. Finally, our calculations with the Navier-Stokes
equations indicate that the Forchheimer model should be valid for low
$Re$ and also for a limited range of high $Re$ numbers, even when
inertial nonlinearities can significantly affect the momentum
transport at the pore scale. However, the magnitude of the deviations
we find at the transition from Darcy to non-Darcy flow suggests a
nonuniversal behavior of the friction factor $f$ with the porosity
$\epsilon$ within this particular region.

\bigskip

\noindent
We thank CNPq, FUNCAP and NSF for support. We also thank three
referees for their constructive criticisms.

\begin{figure}
\caption{ Logarithmic plot showing the dependence of the generalized
friction factor $f$ on the modified Reynolds number $Re'$. The solid
line is the best fit to the Forchheimer equation (4), while the dashed
line is the best fit to Darcy's law of the data at low $Re'$. Please
note that three pairs of $\alpha$ and $\beta$ parameters have been
estimated for each of the three porosity values ($\epsilon=0.7$, $0.8$
and $0.9$). These simulations have been performed with up to five
lattice realizations. The error bars are smaller than the symbols.}
\label{f.1}
\end{figure}
  
\begin{figure}
\caption{ (a) Contour plot of the stream function $\psi$ for low
Reynolds number conditions ($Re=0.0156$). (b) Same as in (a), but for
a Reynolds number 1000 times larger ($Re=15.6$). In both plots, the
values of $\psi$ change by a constant increment between consecutive
streamlines.}
\label{f.2}
\end{figure}

\begin{figure}
\caption{ Dependence of the participation number $\pi$ on the Reynolds
number $Re$ ($\epsilon=0.9$). These simulations have been performed
with ten lattice realizations.}
\label{f.3}
\end{figure}

\begin{figure}
\caption{ (a) Profiles of the velocity magnitude at different
positions in direction $x$ for low Reynolds number conditions
($Re=0.0156$). The realization is the same as in Fig.~2, but the
obstacles have been removed for better visualization.  (b) Same as in
(a), but for a Reynolds number 1000 times larger ($Re=15.6$). The
velocity magnitudes in (a) and (b) have been normalized by the maximum
value calculated for each system.}
\label{f.4}
\end{figure}



\begin{references}

\bibitem{Dul79} F. A. L. Dullien, {\it Porous Media - Fluid Transport
and Pore Structure} (Academic, New York, 1979).

\bibitem{Sah94} M. Sahimi, {\it Applications of Percolation Theory\/}
(Taylor \& Francis, London, 1994); {\it Flow and Transport in Porous
Media and Fractured Rock\/} (VCH, Boston, 1995).

\bibitem{Adl92} P. M. Adler, {\it Porous Media: Geometry and
Transport\/} (Butterworth-Heinemann, Stoneham MA, 1992).

\bibitem{Can90} D. H. Rothman, Geophysics {\bf 53}, 509 (1988);
A. Canceliere, C. Chang, E. Foti, D. H. Rothman, and S. Succi,
Phys. Fluids A {\bf 2}, 2085 (1990).

\bibitem{Kos92} S. Kostek, L. M. Schwartz, and D. L. Johnson,
Phys. Rev. B {\bf 45}, 186 (1992).

\bibitem{Mar92} N. Martys and E. J. Garboczi, Phys. Rev. B {\bf 46},
6080 (1992).

\bibitem{Sch93} L. M. Schwartz, N. Martys, D. P. Bentz,
E. J. Garboczi, and S. Torquato, Phys. Rev. E {\bf 48}, 4584 (1993).

\bibitem{Mar94} N. Martys, S. Torquato, and D. P. Bentz, Phys. Rev. E
{\bf 50}, 403 (1994).

\bibitem{Kop97} A. Koponen, M. Kataja, and J. Timonen, Phys. Rev. E
{\bf 56}, 3319 (1997).

\bibitem{And95} J. S. Andrade, Jr. {\it et al.}, Phys. Rev. E {\bf
51}, 5725 (1995); Phys. Rev. Lett. {\bf 79}, 3901 (1997).

\bibitem{Edw90} D. A. Edwards, M. Shapiro, P. Bar-Yoseph, and
M. Shapira, Phys. Fluids {\bf 2}, 45 (1990).

\bibitem{Koc97} D. R. Koch and A. J. C. Ladd, J. Fluid Mech. {\bf
349}, 31 (1997).

\bibitem{grain} The parameter $d_p$ corresponds to the length of the
side of the square plaquettes taken as obstacles for fluid flow in the
two-dimensional lattice of size $L$. In all simulations $L=1$, which
makes $d_p=1/64$ in length units compatible with the density and
viscosity units.

\bibitem{mesh} We discretize the governing balance equations within
the pore space domain using square grid elements with length equal to
$d_p/4$.  In other words, while the physical system comprises 64x64
elements (solid or fluid cells of size $d_p$), the corresponding
numerical mesh has 256x256 discretization cells. We observe that this
level of refinement generates satisfactory results when compared with
numerical meshes of small resolution.

\bibitem{Pat80} S. V. Patankar, {\it Numerical Heat Transfer and Fluid
Flow\/} (Hemisphere, Washington DC, 1980); The {FLUENT} (trademark of
{FLUENT} Inc.) fluid dynamics analysis package has been used in this
study.

\bibitem{Lee97} S. L. Lee and J. H. Yang, Int. J. Heat Mass Transfer
{\bf 40}, 3149 (1997).

\bibitem{stream} The stream function $\psi$ is defined for
incompressible two-dimensional flows as $u\equiv{\partial\psi/\partial
y}$ and $v\equiv-{\partial\psi/\partial x}$.

\bibitem{Rus94} S. Russ and B. Sapoval, Phys. Rev. Lett. {\bf 73},
1570 (1994); B. Sapoval and S. Russ, Il Nuovo Cimento {\bf 16} D, 1103
(1994).

\bibitem{fsc} To provide a preliminary idea about the finite size
effects on the participation function, we performed simulations with
lattices of size 96x96 (384x384 numerical elements) generated at a
porosity $\epsilon=0.9$ and subjected to low $Re$. The results
averaged over five realizations of the pore structure indicate that
$\pi =0.32 \pm 0.01$, a value which is substantially smaller ($\approx
14 \%$) than the value found for 64x64 systems, $\pi =0.37 \pm 0.01$.

\end{references}
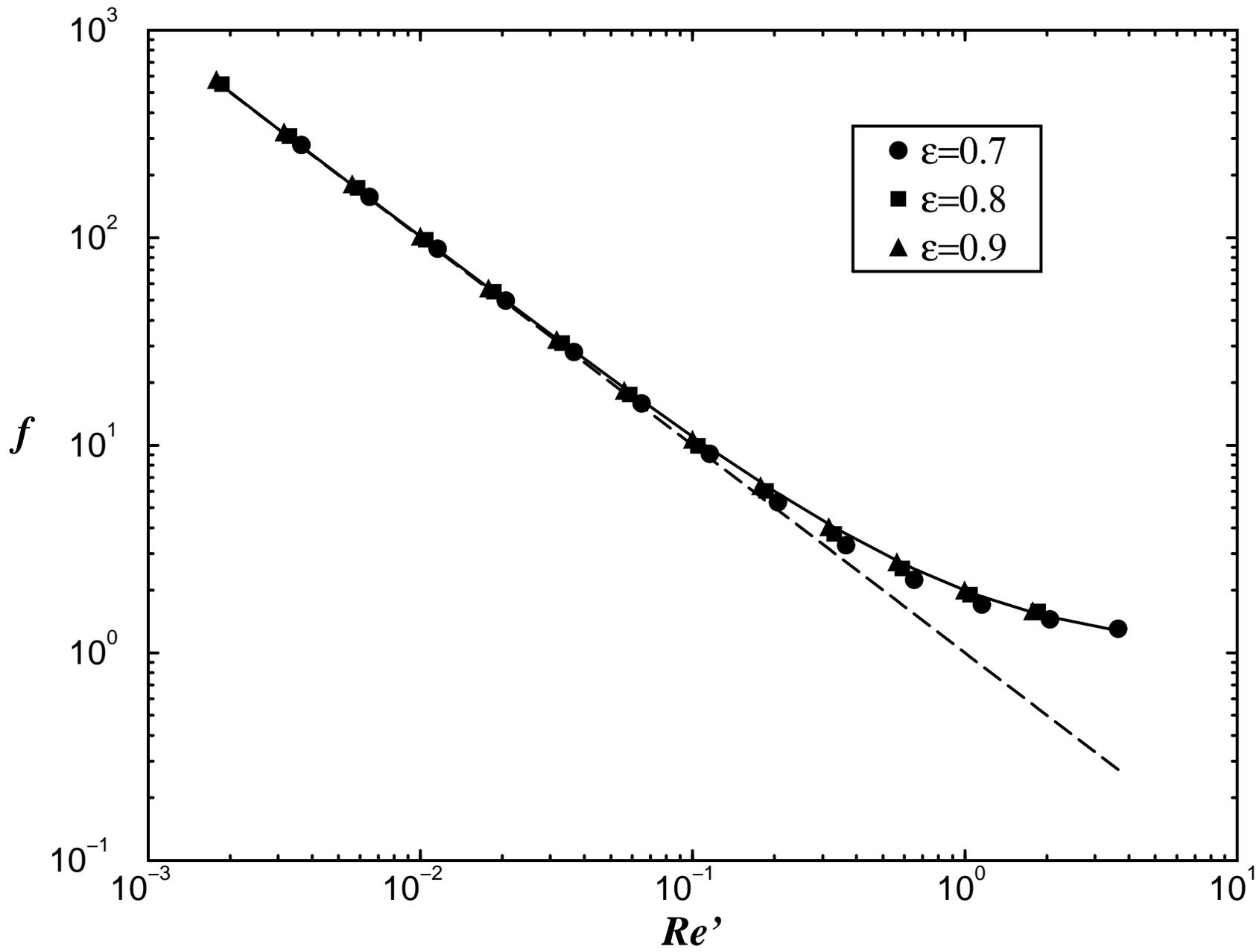
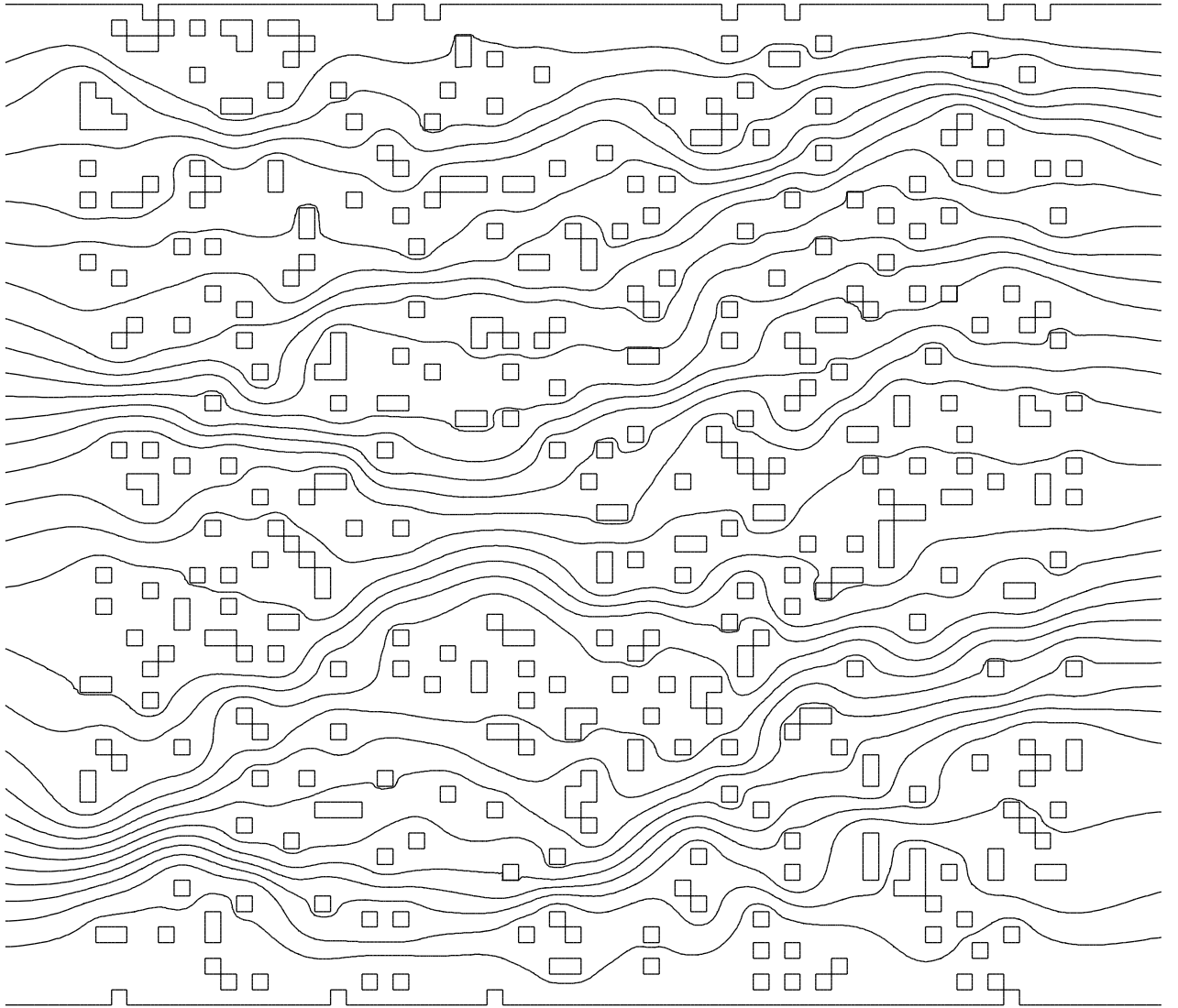
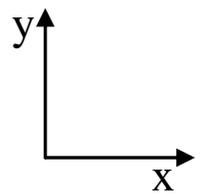
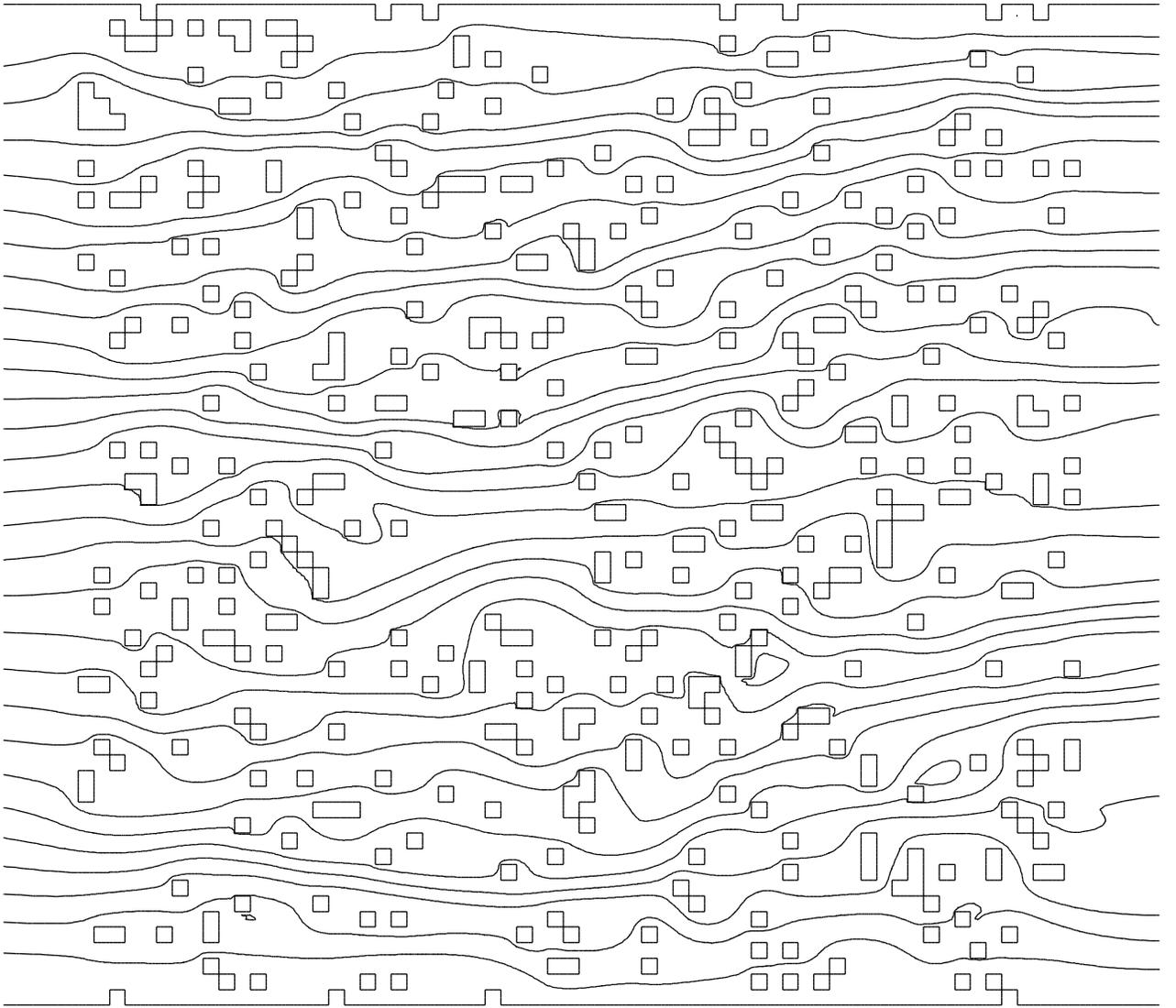
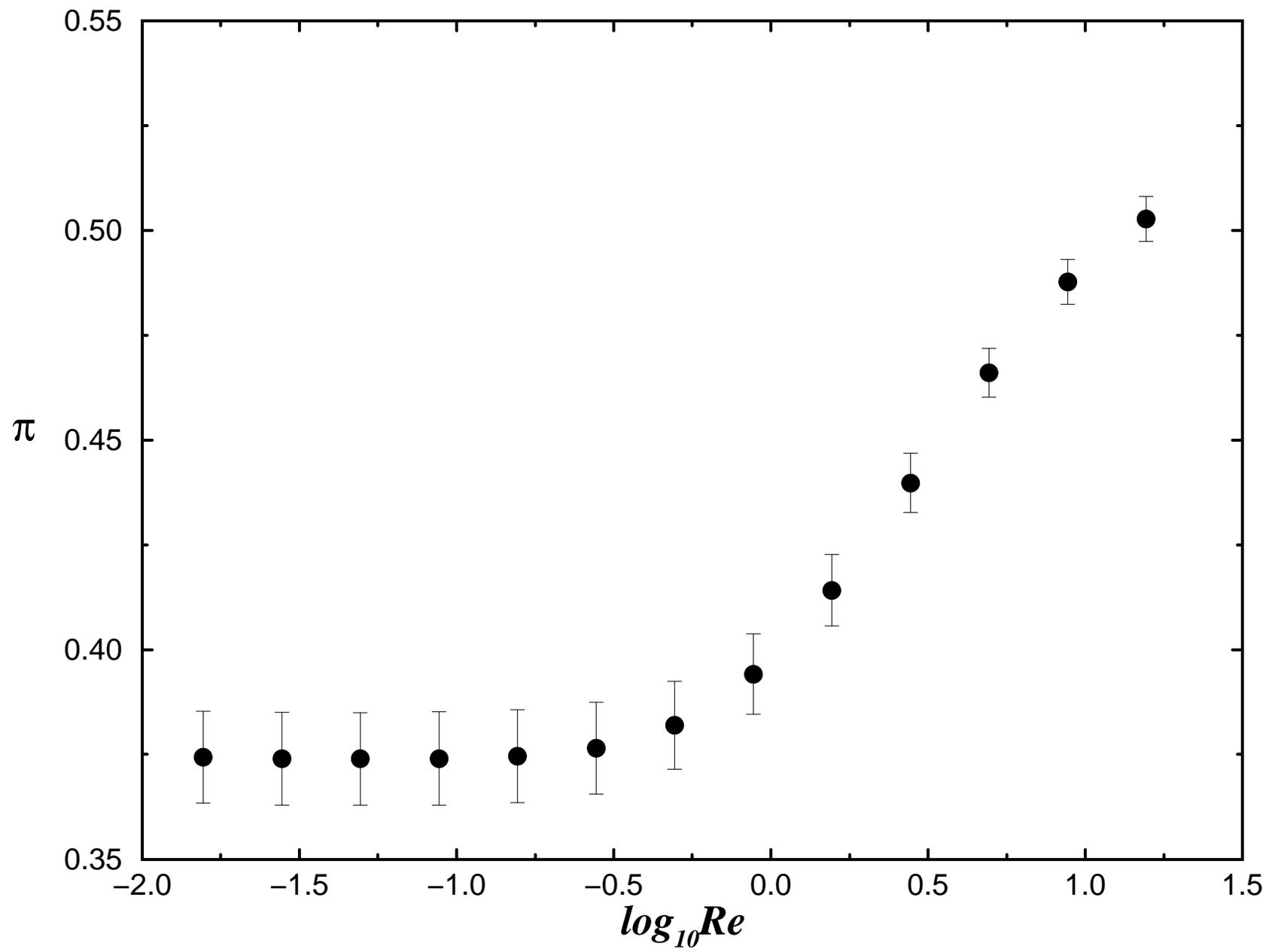
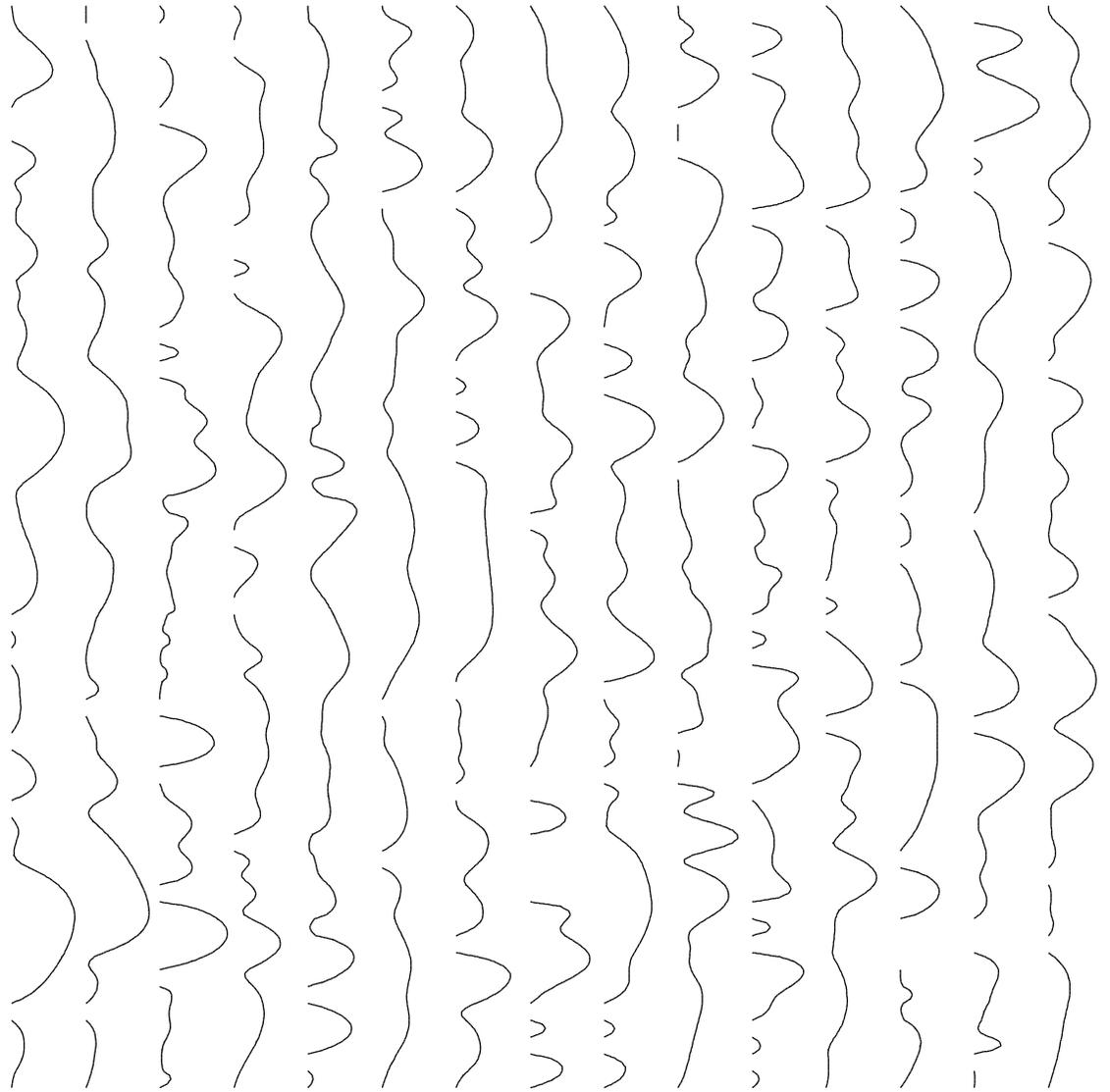
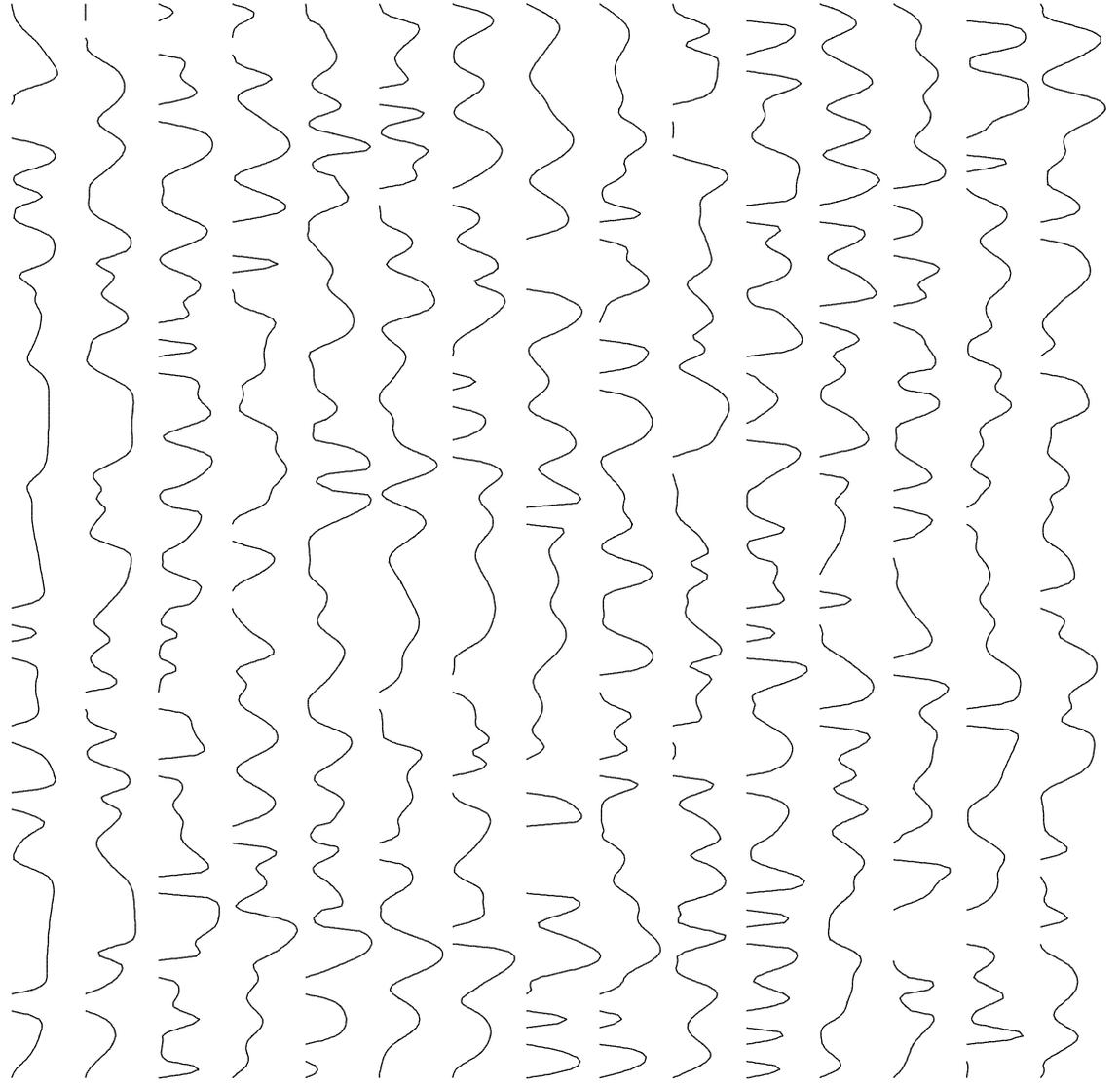
\end{document}